\documentclass[useAMS,usenatbib,usegraphicx]{mn2e}

\title[Ly$\alpha$ and UV emission from high-$z$ GRB hosts]{
  Ly$\alpha$ and UV emission from high-redshift GRB hosts: \\
  To what extent do GRBs trace star formation?\thanks{Based on 
  observations made with the Nordic Optical Telescope, operated on the island 
  of La Palma jointly by Denmark, Finland, Iceland, Norway, and Sweden, in 
  the Spanish Observatorio del Roque de los Muchachos of the Instituto de 
  Astrofisica de Canarias. Based on observations collected at the European
  Southern Observatory Very Large Telescope under programme 071.B-0199(A).}}

\author[P. Jakobsson et al.]{P.~Jakobsson,$^{1,2}$\thanks{E-mail:
pallja@astro.ku.dk} 
G.~Bj\"ornsson,$^{2}$ 
J.~P.~U.~Fynbo,$^{1}$
G.~J\'ohannesson,$^{2}$
J.~Hjorth,$^{1}$
\newauthor
B.~Thomsen,$^{3}$
P.~M\o ller,$^{4}$
D.~Watson,$^{1}$
B.~L.~Jensen,$^{1}$
G.~\"Ostlin,$^{5}$
J.~Gorosabel$^{6,7}$
\newauthor
and E.~H.~Gudmundsson$^{2}$\\
$^{1}$Niels Bohr Institute, University of 
Copenhagen, Juliane Maries Vej 30, DK-2100 Copenhagen, Denmark\\
$^{2}$Science Institute, University of Iceland, Dunhaga 3, 107 Reykjav\'{\i}k, 
Iceland\\
$^{3}$Department of Physics and Astronomy, University of Aarhus, Ny Munkegade, 
DK-8000 \AA rhus C, Denmark\\
$^{4}$European Southern Observatory, Karl-Schwarzschild-Stra\ss e 2, 85748,
Garching bei M\"unchen, Germany\\
$^{5}$Stockholm Observatory, SE-106 91 Stockholm, Sweden\\
$^{6}$IAA-CSIC, P.O. Box 03004, E-18080 Granada, Spain\\
$^{7}$Space Telescope Science Institute, 3700 San Martin Drive, 
Baltimore, MD 21218, USA}

\begin{document}

\date{Accepted 2005 May. Received 2005 May 26; 
in original form 2005 April 11}

\pagerange{\pageref{firstpage}--\pageref{lastpage}} \pubyear{2004}

\maketitle

\label{firstpage}

\begin{abstract}
We report the result of a search for Ly$\alpha$ emission from the host 
galaxies of the gamma-ray bursts (GRBs) 030226 ($z = 1.986$),
021004 ($z = 2.335$) and 020124 ($z = 3.198$). 
We find that the host galaxy of GRB\,021004 is 
an extended (around 8\,kpc) strong Ly$\alpha$ emitter with a restframe 
equivalent width (EW) of $68^{+12}_{-11}$\,\AA, and a star-formation rate
of $10.6 \pm 2.0$\,M$_\odot$\,yr$^{-1}$. We do not detect the hosts 
of GRB\,030226 and GRB\,020124, but the upper limits on their 
Ly$\alpha$ fluxes do not rule out large restframe EWs. In the fields of 
GRB\,021004 and GRB\,030226 we find seven and five other galaxies, 
respectively, with excess emission in the narrow-band filter. These galaxies 
are candidate Ly$\alpha$ emitting galaxies in the environment of the 
host galaxies. We have also compiled a list of all $z \ga 2$ GRB
hosts, and demonstrate that a scenario where they trace star formation in 
an unbiased way is compatible with current observational constraints.
Fitting the $z=3$ luminosity function (LF) under this assumption, results 
in a characteristic luminosity of $R^* = 24.6$ and a faint end slope of 
$\alpha = -1.55$, consistent with the LF measured for Lyman-break galaxies.
\end{abstract}


\section{Introduction}
Ly$\alpha$ photons face a large probability of being absorbed by 
dust particles due to resonant scattering. Therefore, Ly$\alpha$ 
emitting galaxies are often considered to contain 
little or no dust and to have low metallicity \citep{charlot}. However,
a number of local metal-poor galaxies with act\-ive star formation show
little signs of Ly$\alpha$ emission \citep[e.g.][]{mas-hesse,hayes}.
In addition, the luminous dusty SCUBA galaxies often display Ly$\alpha$ 
in emission \citep{chapman}. This indicates that dust and metallicity are 
not the only factors affecting the escape of Ly$\alpha$ photons; 
kinematical and geometrical properties of the interstellar medium may 
also play a major part \citep[e.g.][]{neufeld,gia}. Assuming, however,
the validity of the \citet{charlot} relation between metallicity and 
Ly$\alpha$ emission (see their fig.~8), implies that Ly$\alpha$ imaging is 
a probe of the star-formation rate (SFR), dust content and metallicity 
\par
It is now firmly established that long-duration gamma-ray bursts (GRBs) 
are associated with star formation \cite[e.g.][]{lise} and, at least some,
with core collapse supernovae \citep[e.g.][]{jensN,stanek,malesani,bjarne}.
Ly$\alpha$ observations of GRB host galaxies provide information about 
their SFR, dust content and metallicity, knowledge which is crucial
for our understanding of GRBs, their progenitors and environment. Furthermore,
metallicity is an important parameter in some progenitor models. A high
metallicity produces a strong stellar wind, which in turn leads to mass loss
and loss of angular momentum. In the collapsar model these circumstances make
it extremely problematic to generate a GRB 
\citep[][\hspace{-2.1mm}; see also Petrovic et al. 2005]{mac}; 
a preference for GRB
hosts to be metal-poor is therefore an unambiguous prediction of the collapsar
model (assuming that the progenitor and host follow the same metallicity
distribution). Alternative progenitor models exist where there is no such 
low metallicity preference \citep[e.g.][]{ouyed,fryer}.
\par
There is growing evidence that the majority of GRB host
galaxies, at least at high-$z$, are Ly$\alpha$ emitters \citep{fynbo03a}.
If we define a Ly$\alpha$ emitter as having a Ly$\alpha$ restframe 
equivalent width (EW) larger than 10\,\AA, all of the five possible 
detections are confirmed Ly$\alpha$ emitters: GRB\,971214 at $z = 3.42$ 
\citep{kulkarni98,ahn}, GRB\,000926 at
$z = 2.04$ \citep[][hereafter F02]{fynbo02}, GRB\,011211 at $z=2.14$ 
\citep{fynbo03a}, GRB\,021004 at $z = 2.34$ \citep{moller02} and 
GRB\,030323 at $z=3.37$ \citep{vreeswijk04}. For the latter two bursts, 
the Ly$\alpha$ emission line was detected super\-imposed on the afterglow 
spectrum. In addition, there is an indication of a Ly$\alpha$ emission 
line in the centre of the Ly$\alpha$ absorption trough in the GRB\,030429 
($z = 2.66$) afterglow spectrum, although at a low significance level 
\citep{palli}. As the GRB\,030429 host is faint ($R>26.3$) it would 
imply a relatively high EW.
\par
In this paper we present a study of 
the host galaxies and environments of GRB\,021004 ($z = 2.335$), GRB\,030226 
($z = 1.986$) and GRB\,020124 ($z = 3.198$). The properties of GRB\,021004 
and its afterglow have been discussed in, e.g. \citet{holland}, \cite{nakar}, 
\citet{matheson} and \citet{bersier}. Its host is, like most previously 
studied GRB hosts, a very blue starburst galaxy ($R \approx 24.5$) with no 
evidence for dust \citep{fynbo05}, whereas in the case of GRB\,030226 there 
is no evidence for an underlying host galaxy \citep[$R>26.2$:][]{klose04}. 
GRB\,020124 occurred in a very faint \citep[$R>29.5$:][]{berger} gas-rich
galaxy with evidence for GRB surroundings with low dust content
\citep{jens}. 
\par
We adopt a cosmology where the Hubble constant is 
$H_0 = 70$\,km\,s$^{-1}$\,Mpc$^{-1}$, $\Omega_{\mathrm{m}} = 0.3$ and
$\Omega_{\Lambda} = 0.7$. For these parameters, a redshift 
of (1.986, 2.335, 3.198) corresponds to a luminosity distance
of (15.41, 18.77, 27.45)\,Gpc and a distance modulus of (45.9, 46.4, 47.2). 
One arcsecond is equivalent to (8.38, 8.18, 7.55) proper kiloparsecs, and the
lookback time is (10.2, 10.7, 11.5)\,Gyr.
\section{Observations \& data reduction}
The observations were carried out during one night (two nights were lost
due to bad weather) in October 2003 (GRB\,021004) and four nights in March 
2004 (GRB\,030226) at the 2.56-m Nordic Optical Telescope (NOT) on La Palma, 
using the MOSaic CAmera (MOSCA). The MOSCA detector consists of four
$2048 \times 2048$ CCDs with a pixel scale of $0\farcs217$ ($2 \times 2$ 
binning). The field of GRB\,021004 (GRB\,030226) was imaged in three filters: 
the standard $B$ ($U$) and $R$ filters and a special narrow-band filter 
manufactured by Omega Optical. The narrow-band filters were tuned to 
Ly$\alpha$ at $z = 2.331$ ($\lambda_{\mathrm{cen}} = 4050$\,\AA) and 
$z = 1.986$ ($\lambda_{\mathrm{cen}} = 3630$\,\AA) and had a width of 
$ \Delta \lambda = 60$\,\AA\ (corresponding to a redshift width of 
$\Delta z = 0.049$ for Ly$\alpha$). GRB\,020124 was observed during several 
nights in March and April 2003 with the Very Large Telescope in Chile, using 
the FOcal Reducer and low dispersion Spectrograph (FORS1). The FORS1 
detector consists of a single $2048 \times 2048$ CCD with a pixel scale 
of $0\farcs20$. As this was a Category B service mode 
programme, our observations were not completed, resulting in a total of 
only 5.7\,h of narrow-band imaging\footnote{A total of 18\,hr were granted 
to the programme (to observe the GRB\,020124 host).}. An existing narrow-band 
filter, OIII/6000+52 ($\lambda_{\mathrm{cen}} = 5105$\,\AA) with a width 
of $ \Delta \lambda = 61$\,\AA, was used. For all three fields, the 
observations were carried out more than a year from the onset of the burst. 
An optical afterglow (OA) contribution is therefore negligible.
\par
\begin{table}
  \centering
  \caption{Total exposure time (in hours) and seeing in the combined image 
           for each filter. GRB\,021004 was observed in $B$, while
           GRB\,030226 was observed in $U$. Note that when comparing the
           exposure times, GRB\,020124 was observed with an 8-m telescope.}
  \begin{tabular}{@{}lrrrrrr@{}}
  \hline
GRB & \multicolumn{2}{c}{$R$} & \multicolumn{2}{c}{$B$/$U$} & 
\multicolumn{2}{c}{\hspace{2mm} Narrow-band} \\
    & Exp. & Seeing & Exp. & Seeing & Exp. & Seeing \\
  \hline
021004 & 0.7 & $1\farcs0$ & 1.4 & $0\farcs8$ & 5.0  & $0\farcs9$ \\
030226 & 2.1 & $1\farcs5$ & 7.2 & $1\farcs3$ & 20.0 & $1\farcs3$ \\
020124 & --- & ---        & --- & ---        & 5.7  & $0\farcs8$ \\
  \hline
  \end{tabular}
\label{exp.tab}
\end{table}
The individual exposures were bias-subtracted and flat-field corrected using 
standard techniques. The individual reduced images were combined using a 
$\sigma$-clipping routine optimised for faint sources \citep[see details 
in][]{moller03}. The total integration times along with the seeing in the 
combined images are given in Table~\ref{exp.tab}. Narrow-band flux calibration 
was performed using observations of the spectrophotometric standard stars 
BD+28, GD71 and Hz44. The broadband images were calibrated using the 
\citet{henden02,henden03} secondary standards. The transformations given 
in \citet{fuk} were finally used to bring the observations onto the 
AB-system. Conditions were not photometric when the NOT data were obtained, 
resulting in the narrow-band zero\-points being affected by clouds. Hence, 
they had to be adjusted as detailed in Sect.~\ref{grb021004.sec}. This 
effect has not been taken into account in the uncertainties quoted in the 
paper.
\section{Results}
\label{results.sec}
We have used the same methods for photometry and selection of Ly$\alpha$
emitting galaxy candidates as those described in F02. In the 
following we will use the acronym LEGOs for Ly$\alpha$ Emitting 
Galaxy-building Objects, introduced by \citet{moller01}.
\subsection{GRB\,021004}
\label{grb021004.sec}
\begin{figure}
   \centering
\resizebox{\hsize}{!}{\includegraphics[bb=12 0 493 663,clip]{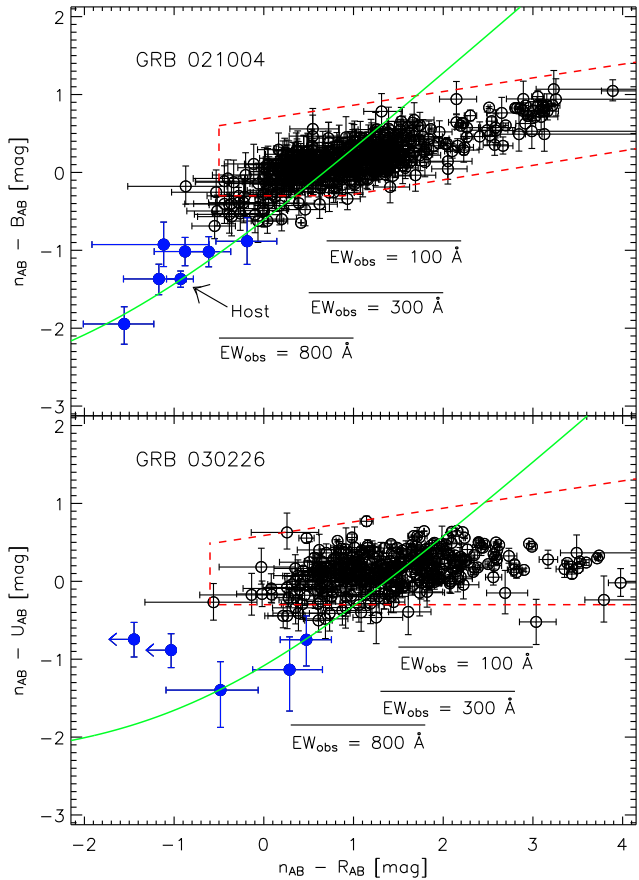}}
   \caption{Colour-colour diagram for all objects detected at $S/N > 5$ in 
            the narrow-band image, in the fields of GRB\,021004 (top) and 
            GRB\,030226 (bottom). The dashed box encloses an area where all 
            the model galaxy spectra are located (see main text for details). 
            As expected, most objects have colours consistent with being in 
            this box. However, a number of objects, including the GRB\,021004 
            host, are located in the lower left part of the dia\-grams. The 
            filled circles indicate LEGO candidates. The solid curve 
            corresponds to objects having the same broadband colours, but 
            various amounts of absorption (upper part) or emission (lower 
            part) in the narrow-band filter.}
\label{colour.fig}
\end{figure}

In the upper part of Fig.~\ref{colour.fig}, the 
$n_{\mathrm{AB}}$$-$$B_{\mathrm{AB}}$ versus
$n_{\mathrm{AB}}$$-$$R_{\mathrm{AB}}$ colour-colour diagram is plotted
for the detected objects in the GRB\,021004 field. The colours have
not been corrected for Galactic extinction; such a correction would
be negligible for $n_{\mathrm{AB}}$$-$$B_{\mathrm{AB}}$ and only shift
objects slightly to the left (thus not affecting the number of LEGOs
detected). In order to constrain
where objects with no special features in the narrow-band filter fall
in the diagram, we have calculated colours based on the synthetic 
galaxy spectra without Ly$\alpha$ emission provided by \citet{bruzual}. We 
have used models with ages ranging from a few Myr to 15 Gyr, 
with $0 < z < 3$. The box in Fig.~\ref{colour.fig} indicates the area where 
these model galaxies are located. 
\par
As mentioned above the narrow-band zeropoints for the NOT data were 
affected by cirrus. They were adjusted by requiring that the colours of 
objects detected at fairly high signal-to-noise ratio ($S/N$) would fall 
on the locus of the model galaxies. The resulting change in the 
narrow-band zeropoint was approximately 25\%, with an uncertainty of 
approximately 5\%.
\par
\begin{table}
  \centering
  \caption{Photometric properties of the seven LEGO candidates in
           the field of GRB\,021004. The GRB\,021004 host galaxy
           is denoted by S1004\_5.}           
  \begin{tabular}{@{}lccrr@{}}
  \hline
Object           & $R_{\mathrm{AB}}$    & $B_{\mathrm{AB}}$ &
$f$(Ly$\alpha$)  & SFR(Ly$\alpha$)      \\           
                 &                      &                   &
[$10^{-17}$\,erg &                      \\
                 & [mag]                & [mag]             &
s$^{-1}$\,cm$^{-2}$] & [M$_\odot$\,yr$^{-1}$] \\

\hline

S1004\_1 & $23.83^{+0.22}_{-0.19}$ & $24.41^{+0.16}_{-0.14}$ & $11.8 \pm 3.6$
& $5.0 \pm 1.5$ \\

S1004\_2 & $24.22^{+0.24}_{-0.19}$ & $25.23^{+0.25}_{-0.20}$ & $3.2 \pm 1.7$
& $1.4 \pm 0.7$ \\

S1004\_3 & $24.98^{+0.52}_{-0.35}$ & $25.41^{+0.29}_{-0.23}$ & $10.2 \pm 3.2$
& $4.3 \pm 1.4$ \\

S1004\_4 & $25.49^{+1.23}_{-0.56}$ & $25.00^{+0.21}_{-0.18}$ & $7.2 \pm 2.4$
& $3.1 \pm 1.0$ \\

S1004\_5 & $24.18^{+0.33}_{-0.25}$ & $24.43^{+0.16}_{-0.14}$ & $25.1 \pm 4.8$
& $10.6 \pm 2.0$ \\

S1004\_6 & $24.92^{+0.48}_{-0.33}$ & $25.43^{+0.30}_{-0.23}$ & $11.3 \pm 4.0$
& $4.8 \pm 1.7$ \\

S1004\_7 & $24.22^{+0.18}_{-0.15}$ & $25.42^{+0.23}_{-0.19}$ & $3.2 \pm 1.7$
& $1.4 \pm 0.7$ \\

  \hline
  \end{tabular}
\label{flux1.tab}
\end{table}
\begin{table}
  \centering
  \caption{Photometric properties of the five LEGO candidates in
           the field of GRB\,030226.}
  \begin{tabular}{@{}lccrr@{}}
  \hline
Object           & $R_{\mathrm{AB}}$    & $U_{\mathrm{AB}}$ &
$f$(Ly$\alpha$)  & SFR(Ly$\alpha$)      \\           
                 &                      &                   &
[$10^{-17}$\,erg &                      \\
                 & [mag]                & [mag]             &
s$^{-1}$\,cm$^{-2}$] & [M$_\odot$\,yr$^{-1}$] \\

  \hline
S0226\_1 & $24.27^{+0.15}_{-0.13}$ & $24.91^{+0.10}_{-0.09}$ & $2.3 \pm 1.2$
& $0.7 \pm 0.3$ \\

S0226\_2 & $24.62^{+0.17}_{-0.15}$ & $25.32^{+0.11}_{-0.10}$ & $3.5 \pm 1.8$
& $1.0 \pm 0.5$ \\

S0226\_3 & $24.81^{+0.21}_{-0.18}$ & $25.97^{+0.21}_{-0.18}$ & $2.7 \pm 1.6$
& $0.8 \pm 0.5$ \\

S0226\_4 & $>$24.9                 & $25.45^{+0.27}_{-0.22}$ & $4.9 \pm 1.9$
& $1.4 \pm 0.5$ \\

S0226\_5 & $>$24.9                 & $25.16^{+0.11}_{-0.10}$ & $2.4 \pm 1.0$
& $0.7 \pm 0.3$ \\

  \hline
  \end{tabular}
\label{flux2.tab}
\end{table}
\begin{figure*}
   \centering
   \includegraphics[width=10.0cm]{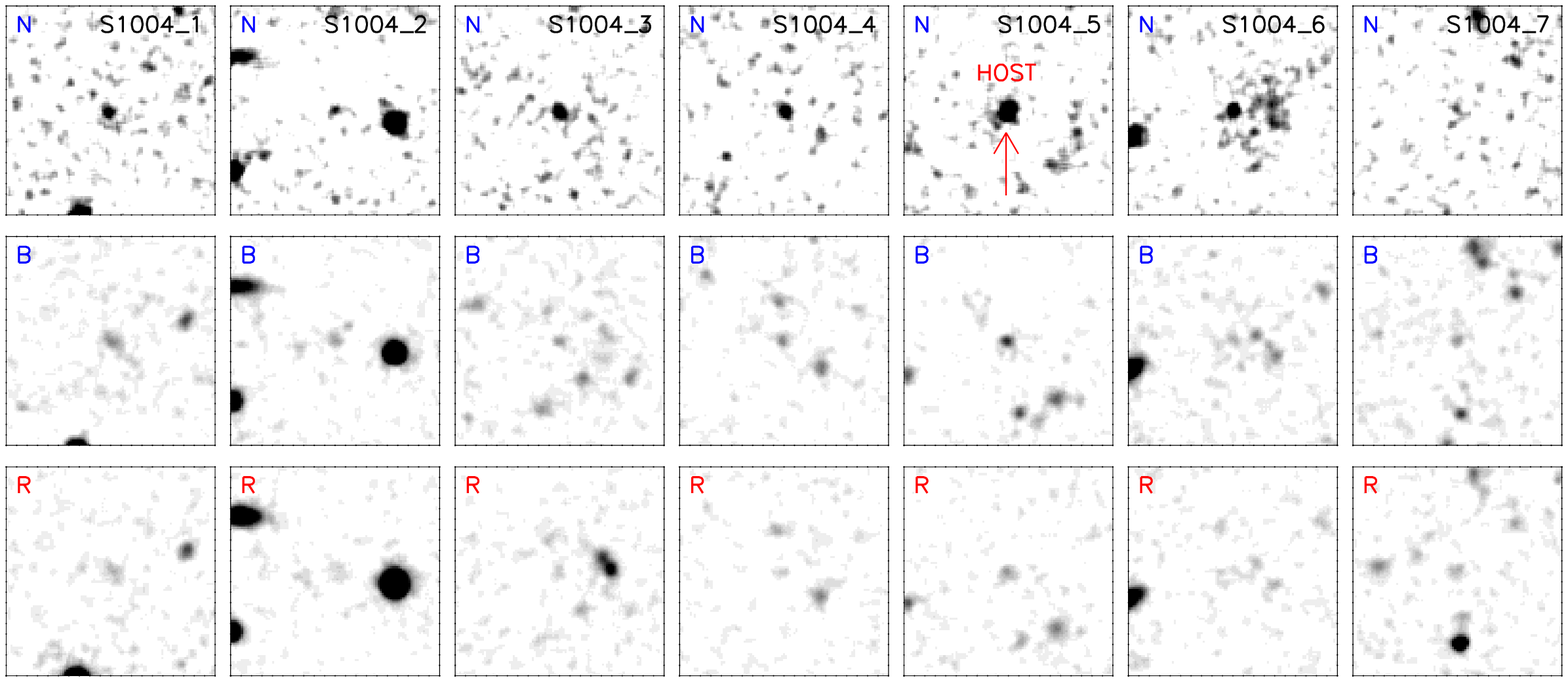}
   \includegraphics[width=7.4cm,bb=-20 0 488 299,clip]{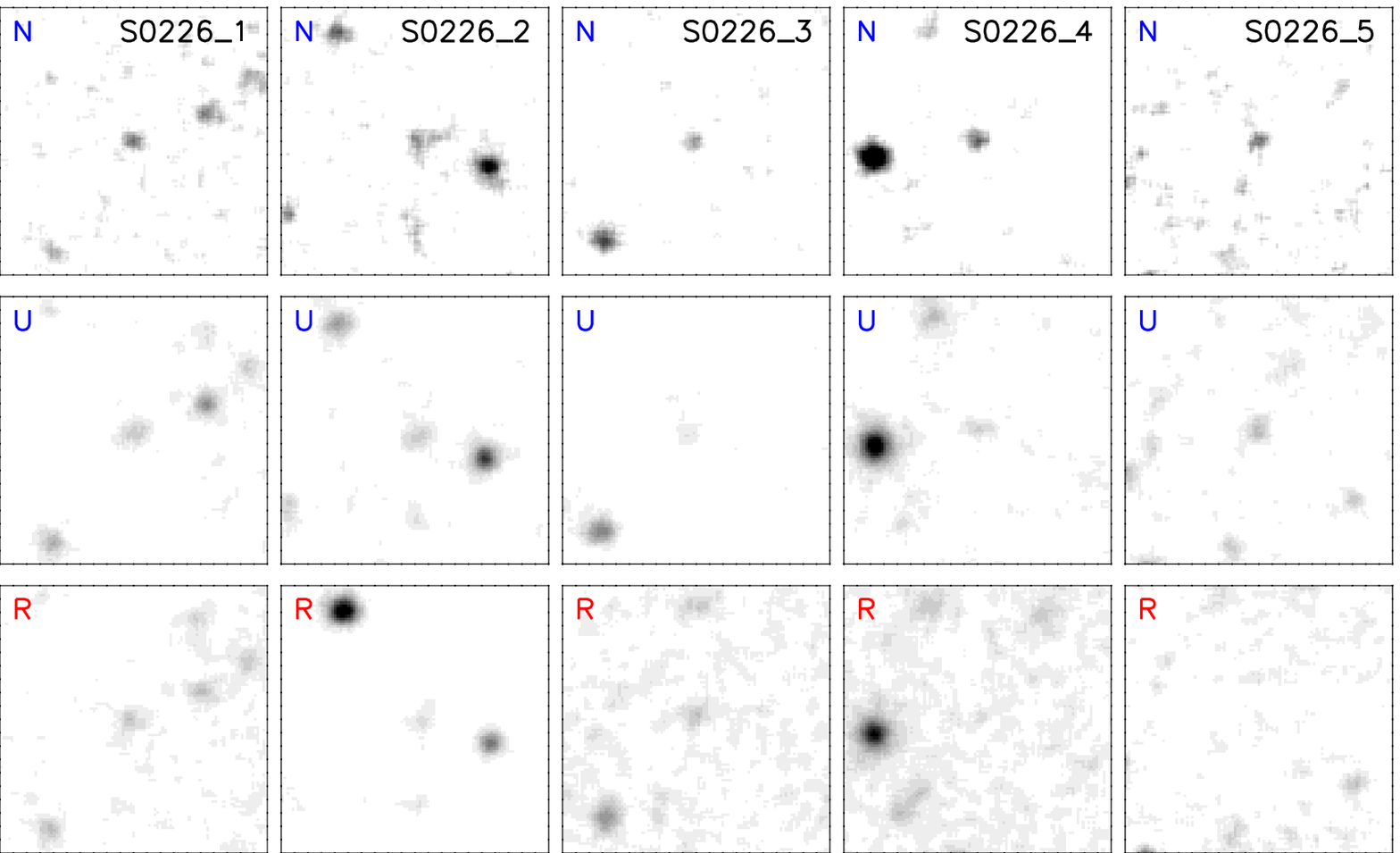}
   \caption{A $10\arcsec \times 10\arcsec$ section around each of the 
            LEGOs in the field of view of GRB\,021004 (left) and 
            GRB\,030226 (right). For each candidate we show a sub-image 
            from all three filters: narrow-band (top row), $B$-band 
            or $U$-band (middle row) and $R$-band (bottom row). North 
            is up and east is to the left. The GRB\,021004 host is 
            named S1004\_5.}
\label{cand.fig}
\end{figure*}
LEGOs will fall in the lower left corner of the diagram, due
to excess emission in the narrow-band filter. We select as LEGO 
candidates objects detected at more than 5$\sigma$ in the narrow-band
image and with a colour $n_{\mathrm{AB}}$$-$$B_{\mathrm{AB}} < -0.7$.
This corresponds to an observed EW of about 67\,\AA, or 20\,\AA\ in 
the restframe for Ly$\alpha$ at $z = 2.335$. We find seven candidates in
the field, including the host galaxy. Images of these LEGOs are shown in 
the left part of Fig.~\ref{cand.fig} and their photometric properties
based on the total magnitudes (\texttt{mag\_auto}) from SExtractor
\citep{sex} are given in Table~\ref{flux1.tab}. The Ly$\alpha$ line
flux is calculated by dividing the flux in the narrow-band filter with
$1 + \Delta \lambda / \textrm{EW}_{\mathrm{obs}}$. We also derive the 
SFR for the LEGOs, assuming that a Ly$\alpha$ luminosity of 
$10^{42}$\,erg\,s$^{-1}$ corresponds to a SFR of 
$1\,\textrm{M}_{\odot}\,\textrm{yr}^{-1}$ \citep{kennicutt,hu}. These 
SFRs should be considered as lower limits, since the estimates might be
affected by extinction.
\par
\begin{figure}
   \centering
   \resizebox{\hsize}{!}{\includegraphics{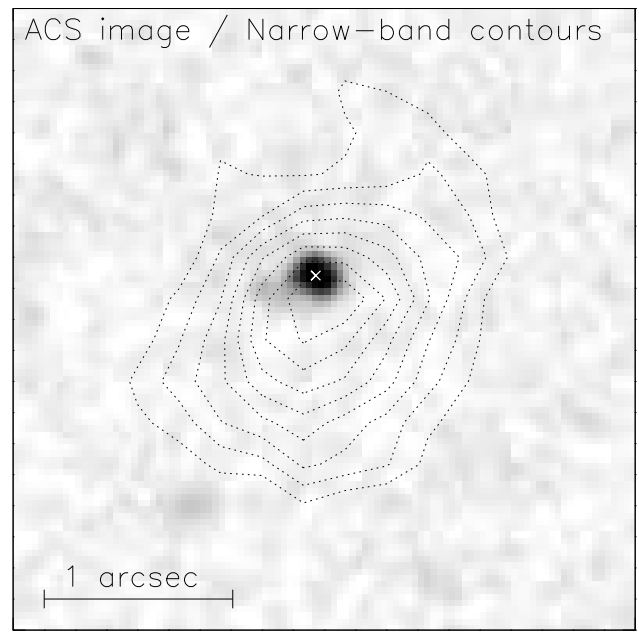}}
   \caption{A $3\farcs2 \times 3\farcs2$ section of an HST/ACS/WFC image 
            centered on the host galaxy of GRB\,021004. It was obtained on
            2003 May 31, approximately 239 days after the burst. 
            The contours show the Ly$\alpha$ emission based on the
            narrow-band observations (with a seeing of $0\farcs9$). The 
            GRB (position marked with a cross) went off near the centre 
            of the galaxy. North is up and east is to the left.}
\label{1004_hst.fig}
\end{figure}

The host galaxy of GRB\,021004 is detected in all bands and is a 
Ly$\alpha$ emitter with a restframe EW of $68^{+12}_{-11}$\,\AA. Deep 
HST/ACS images of the host have been reported by \citet{fynbo05}. 
These images show that the host is dominated by a single component with 
the $fwhm$ extending over $0\farcs12$ or roughly 1\,kpc at $z = 2.335$. 
In Fig.~\ref{1004_hst.fig} the contours of the Ly$\alpha$ emission are
overplotted on an HST/ACS/WFC image, obtained 239 days after the 
burst. The Ly$\alpha$ weighted centroid is located very close to
(and is consistent with) the centre of the galaxy, where the GRB 
originated. The extent of the Ly$\alpha$ emission is much larger
than that of the continu\-um emission (about $1\farcs0$ after correcting
for the seeing). A similar difference between continuum and Ly$\alpha$
emitting regions is seen both for other high-$z$ Ly$\alpha$
emitters \citep[e.g.][]{fynbo03b} and for local starburst galaxies 
\citep{hayes}.
\subsection{GRB\,030226}
The host galaxy of GRB\,030226 is not detected in any of the bands. We
can set an upper limit ($5 \sigma$) of \mbox{$n_{\mathrm{AB}} \approx 25.8$}, 
corresponding to an observed emission line flux of $2.4 \times 
10^{-17}$\,erg\,s$^{-1}$\,cm$^{-2}$. We calculate the broadband limiting
magnitudes ($2 \sigma$) in a circular aperture with a radius equal to the 
seeing, resulting in $R_{\mathrm{AB}} \approx 24.9$ and $U_{\mathrm{AB}} 
\approx 26.2$. In Fig.~\ref{hosts.fig} we show a $10\arcsec \times 
10\arcsec$ region containing the position of the afterglow from a weighted 
sum of the combined Ly$\alpha$ and $U$-band images. The weight is calculated 
as the variance of the background after scaling to the same flux level. We
detect no significant emission from the host above an estimated $2 \sigma$ 
limit of $U_{\mathrm{AB}} \approx 26.5$.
\par
In the lower part of Fig.~\ref{colour.fig}, the 
$n_{\mathrm{AB}}$$-$$U_{\mathrm{AB}}$ versus
$n_{\mathrm{AB}}$$-$$R_{\mathrm{AB}}$ colour-colour diagram is plotted
for the detected objects in the GRB\,030226 field. We have followed a
similar procedure as in Sect.~\ref{grb021004.sec} to select the LEGOs, 
resulting in the detection of five candidates. Here the colour criteria
is $n_{\mathrm{AB}}$$-$$U_{\mathrm{AB}} < -0.6$, corresponding to an 
observed EW of about 60\,\AA, or 20\,\AA\ in the restframe for 
Ly$\alpha$ at $z = 1.986$. The LEGO images are shown in the right part of
Fig.~\ref{cand.fig} and their photometric properties are given 
in Table~\ref{flux2.tab}.
\subsection{GRB\,020124}
Lacking broadband observations for the GRB\,020124 field, we are unable
to identify LEGO candidates. The host galaxy is not detected above a limit
of $1.5 \times 10^{-17}$\,erg\,s$^{-1}$\,cm$^{-2}$ ($5 \sigma$) in the
narrow-band image. Since the host is very faint \citep{berger}, this 
non-detection does not exclude a large Ly$\alpha$ EW (see below). In 
Fig.~\ref{hosts.fig} we show a $10\arcsec \times 10\arcsec$ region 
centred on the position of the afterglow.
\subsection{Foreground Interlopers}
Low-redshift galaxies with other emission lines in the narrow-band
filter can mimic the high-redshift Ly$\alpha$ candidates we are aiming at
\citep[e.g.][]{stern}. For the GRB\,021004 host, the redshift is known
from spectroscopy \citep{moller02}. For the remaining LEGO candidates, 
other emission line sources than Ly$\alpha$ are possible. We can exclude
foreground galaxies with [\mbox{O\,{\sc ii}}] $\lambda$3727 in the
GRB\,021004 narrow-band filter, since such objects would have to be at
a very low redshift ($z = 0.09$) and hence be much brighter than our
candidates. Other possible explanations are \mbox{C\,{\sc iv}} at 
$z = 1.34$ (GRB\,030226) and $z = 1.61$ (GRB\,021004), and 
\mbox{Mg\,{\sc ii}} at $z = 0.30$ (GRB\,030226) and $z = 0.45$ 
(GRB\,021004). The presence of strong \mbox{C\,{\sc iv}} and 
\mbox{Mg\,{\sc ii}} emission requires an underlying active galactic
nucleus (AGN). We have carried out a similar calculation as in 
F02 to show that the probability, that the expected number
of AGNs in our field of view fall in the aforementioned small volumes,
is of the order of 1\%. Therefore, significant \mbox{Mg\,{\sc ii}} and
\mbox{C\,{\sc iv}} contamination due to AGNs is not expected.
\begin{table*}
\begin{minipage}{159mm}
  \caption{A list of all GRBs with $z > 1.9$ known to date (May 2005). 
           The detection method indicates how Ly$\alpha$ in emission was
           detected. The host magnitudes are given in the Cousins/Johnson
           photometric system and have been corrected for Galactic
           extinction \citep{{schlegel}}. The restframe wavelength is
           obtained by dividing the observed wavelength by ($1 + z$). 
           References are given in order for: redshift, 
           Ly$\alpha$ EW$_{\mathrm{rest}}$ and host magnitude. 
          (1) \citet{kulkarni98}; (2) S.-H. Ahn (private communication);
          (3) \citet{odewahn};    (4) \citet{andersen};
          (5) This work;          (6) \citet{brian};
          (7) F02;                (8) \citet{fv};
          (9) \citet{castro};     (10) \citet{fruchter};
          (11) \citet{fynbo03a};  (12) \citet{p03};
          (13) \citet{jens};      (14) \citet{berger};
          (15) \citet{moller02};  (16) \citet{fynbo05};   
          (17) \citet{klose04};   (18) \citet{vreeswijk04};
          (19) \citet{palli};     (20) \citet{kelson};    
          (21) \citet{javier};    (22) \citet{fynbo0319}; 
          (23) \citet{fynbo0401}; (24) \citet{prochaska};
          (25) \citet{brad};      (26) \citet{edo427};
          (27) \citet{chapmanR}.
 }
  \begin{tabular}{@{}lcclrcrl@{}}
  \hline
   GRB  & $z$  & Ly$\alpha$ EW$_{\mathrm{rest}}$ & Detection method
        & Host mag & Restframe & Host mag & References \\
        &      & [\AA] &   &   & wavelength [\AA]   & ($z = 3$) & \\
  \hline
971214  & 3.42 & $\sim$14         & Ly$\alpha$ line in host spectrum 
& $R = 26.5$ & 1490 & 26.4 & (1) (2) (3) \\

000131  & 4.50 & ---              &                
& $I = 26.4$ & 1465 & 26.2 & (4) (5)     \\

000301C & 2.04 & $\la$150         &                
& $R = 27.9$ & 2165 & 28.1 & (6) (7) (8) \\

000926  & 2.04 & $71^{+20}_{-15}$ & Narrow-band Ly$\alpha$ imaging        
& $U = 23.9$ & 1200 & 24.1 & (9) (7) (7)  \\

011211  & 2.14 & $21^{+11}_{-8}$  & Narrow-band Ly$\alpha$ imaging       
& $R = 25.0$ & 2095 & 25.2 & (10) (11) (12) \\

020124  & 3.20 & ---              &                
& $R > 29.4$ & 1565 & $>$29.4 & (13) (14)   \\

021004  & 2.34 & $68^{+12}_{-11}$ & Narrow-band Ly$\alpha$ imaging       
& $B = 24.3$ & 1330 & 24.4 & (15) (5) (16) \\

030226  & 1.99 & ---      &                
& $U > 25.7$ & 1220 & $>$25.9 & (17) (5) \\

030323  & 3.37 & $\sim$145       & Ly$\alpha$ line in OA spectrum 
& $V = 27.9$ & 1260 & 27.8 & (18) (18) (18) \\

030429  & 2.66 & ---      &                
& $R > 26.1$ & 1800 & $>$26.2 & (19) (19) \\

050315  & 1.95 & ---      &                
& $R = 23.8$ & 2230 & 24.0 & (20) (21) \\

050319  & 3.24 & ---      &                
& $R > 25.0$ & 1550 & $>$25.0 & (22) (5) \\

050401  & 2.90 & ---      &                
& $R > 25.3$ & 1685 & $>$25.3 & (23) (5) \\

050502A & 3.79 & ---      &                
& $R > 23.5$ & 1305 & $>$23.4 & (24) (25) \\

050505  & 4.27 & ---      &                
& $R > 21.5$ & 1250 & $>$21.3 & (26) (27) \\

  \hline
  \end{tabular}
\label{list.tab}
\end{minipage}
\end{table*}
\section{Discussion}
\subsection{LEGOs}
We compare the properties of the LEGO candidates in the fields of 
GRB\,021004 and GRB\,030226 to properties of $z \approx 2$ galaxies found 
in other surveys. The survey parameters (including flux limits) of 
\citet{kurk} and \citet{pentericci}, who have detected and 
spectroscopically confirmed 15 \mbox{LEGOs} (including the radio
galaxy) in the field around PKS\,1138$-$262 at $z = 2.156$, are similar to 
ours. Their inferred number of LEGOs per arcmin$^2$ per unit redshift 
is $7.8 \pm 2.0$. The 
corresponding number for the fields of GRB\,000301C and GRB\,000926 
(F02) is $8.3\pm 2.0$. In addition, \citet{fynbo03a} find 
$\sim$4.5 LEGOs per arcmin$^2$ per unit redshift in the field of 
GRB\,011211.
\par
We observe 12 LEGO candidates in two $\sim$50\,arcmin$^2$ fields with a
filter spanning $\Delta z = 0.049$. This corresponds to $\sim$2.5 LEGOs per 
arcmin$^2$ per unit redshift. We note that the MOSCA field of view is around 
60\,arcmin$^2$, but the combined effect of the relatively large chip
gaps ($\sim$8\arcsec) and image dithering resulted in a loss of 
approximately 10\,arcmin$^2$. The LEGO mean density for the five GRB
fields observed thus far is about 5.6, consistent with that 
found around PKS\,1138$-$262. Unfortunately, there are currently no similar 
(in redshift and depth) surveys for Ly$\alpha$ emitters in blank fields. 
Thus, the number of LEGOs representing the mean density at $z \approx 2$ 
is uncertain. But there is some evidence that GRB host galaxies do not 
reside in high galaxy density environments 
\citep[][\hspace{-2.35mm}; Foley et al. in preparation]{bornancini}.
\subsection{Ly$\alpha$ Emission from the GRB Hosts}
This study confirms what is already known about GRB\,021004, namely that
its host is a strong Ly$\alpha$ emitter with a restframe EW of 
$68^{+12}_{-11}$\,\AA, consistent with the value of $69 \pm 14$\,\AA\
estimated by \citet{fynbo05} from a ground-based spectrum. The lack of 
Ly$\alpha$ emission from the hosts of GRB\,030226 and GRB\,020124 in our 
deep narrow-band images does not rule out large restframe EWs. Assuming
a GRB\,020124 host magnitude of $R_{\mathrm{AB}} \approx 29.5$ 
\citep[consistent with the deep limit from][]{berger}, a restframe 
Ly$\alpha$ EW of $\sim$500\,\AA\ would have remained undetectable. The 
corresponding value for GRB\,030226, assuming a host magnitude of 
$U_{\mathrm{AB}} \approx 26.5$, is $\sim$40\,\AA. Fainter host magnitudes
would result in even higher restframe EW limits.
\subsection{Are GRBs Biased Tracers of Star Formation?}
All current evidence is consistent with the conjecture that the host 
galaxies of GRBs, at least at high redshifts ($z \ga 2$), are Ly$\alpha$ 
emitters. An explanation could be an OA selection bias against dusty hosts
\citep[e.g.][and references therein]{tanvir}. However, the fraction of 
truly dark bursts is perhaps as small as 10\% 
\citep[e.g.][]{fynbo01,berger,lamb,palliApJL,palliApJ}. This suggests that
GRB-selected galaxies should not be biased against very dusty systems.
\par
A low metallicity preference for GRB progenitors, at least at high-$z$,
is a possible interpretation of our results \citep{fynbo03a}. In this case 
GRBs could be \emph{biased} tracers of star formation \citep[e.g.][]{le}. 
It is therefore of interest to consider whether our results are 
consistent with the hypothesis that GRBs are \emph{unbiased} tracers of star 
formation between approximately $2 \la z \la 4$, where the majority of known 
high-$z$ GRBs are located, and where the peak of the star-formation activity
is situated.
\par
In order to test this hypothesis we have collected all high-$z$ ($z \ga 2$) 
GRB host galaxies reported in the literature, together with their redshifts, 
Ly$\alpha$ restframe EW if available, host magnitudes and host magnitudes 
redshifted to $z = 3$, assuming a flat $F_{\nu}$ spectrum 
(Table~\ref{list.tab}). Assuming that GRBs trace UV luminosity we fit the 
$z = 3$ magnitudes to the \citet{schechter} luminosity function (LF), 
including the upper limits as described in Appendix~\ref{fit.app}. We find 
a characteristic magnitude of $24.6^{+0.6}_{-0.7}$\,mag and a faint end 
slope of $\alpha = -1.55^{+0.24}_{-0.16}$ (the 1$\sigma$ contour is
shown in Fig.~\ref{maxlik.fig}). We compare this to the $R$-band LF, 
corresponding to the restframe UV LF, for Lyman-break galaxies (LBGs) at 
$z = 3$. The LBG LF has a characteristic magnitude of $24.54 \pm 0.14$\,mag 
and a faint end slope of $\alpha = -1.57\pm0.14$ \citep{adelberger00}. Hence, 
there is consistency between the properties of the GRB host and LBG LFs, 
implying that GRBs are consistent with tracing the UV luminosity. Only 
$\sim$33\% of the $R<25.5$ LBGs are Ly$\alpha$ emitters with a restframe 
EW larger than 10\,\AA\ \citep{shapley}, but if this fraction is higher at 
fainter magnitudes this could be consistent with the high fraction of 
Ly$\alpha$ emitters among GRB hosts.
\par
\begin{figure}
   \centering
   \resizebox{\hsize}{!}{\includegraphics{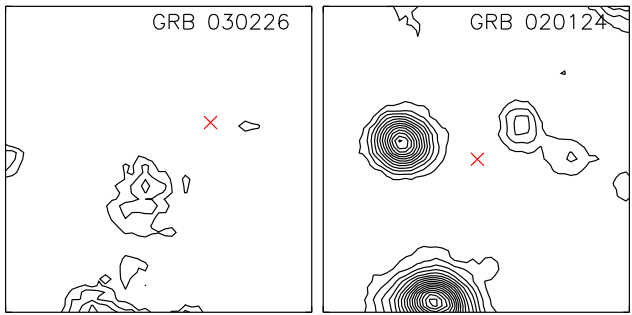}}
   \caption{The location of GRB\,030226 (left) and GRB\,020124 (right)
            as imaged in the combined Ly$\alpha + U$-band (left) and 
            Ly$\alpha$ (right). The size of the images is $10\arcsec \times 
            10\arcsec$. The position of the optical afterglows (OAs) is 
            indicated with a cross. In both panels, the OA positional 
            uncertainty is smaller than the extent of the cross. No 
            significant emission at the burst positions is detected. 
            North is up and east is to the left.}
\label{hosts.fig}
\end{figure}
The next question is whether UV luminosity on average is proportional to
the SFR. Locally, such a correlation has been established \citep{kennicutt},
but it is known that for some starburst galaxies, most of the 
UV emission is absorbed and re-radiated by dust \citep[e.g.][]{chapman}.
However, \citet{chary} argue that the contribution to the total SFR
density at $z=3$ from dust-enshrouded ultra-luminous IR galaxies (ULIRGs) is
most likely less than 30\%. Therefore it is a good first approximation to 
assume that the UV luminosity, for the bulk of the star-forming galaxies, 
is proportional to the SFR at $z=3$. In conclusion, the available data do 
not exclude GRBs as star-formation tracers, but additional observations are 
clearly required to settle the issue.
\par
If the mass-metallicity relation for galaxies \citep{tremonti} was already 
in place at $z \ga 2$ \citep{moller04,ledoux}, a natural interpretation of 
our Ly$\alpha$ results, is that GRB hosts 
predominantly are low-mass systems. 
It has been argued from 
redshift surveys that there is downsizing in galaxy formation, i.e. that 
the most massive galaxies form first and galaxy formation proceeds from 
larger to smaller mass scales \citep[e.g.][]{juneau}. This scenario, 
seemingly at odds with the hierarchical galaxy formation scenario, is not 
supported by the properties of high-$z$ GRB host galaxies if they trace 
star formation: under this assumption, low-mass galaxies dominate the 
integrated star-formation density at high-$z$. The evidence from redshift 
surveys then only implies that star-formation activity in high-mass galaxies 
dies out faster than in low-mass galaxies, which seem to dominate at all 
redshifts.
\begin{figure}
   \centering
   \resizebox{\hsize}{!}{\includegraphics[bb=75 375 539 696,clip]{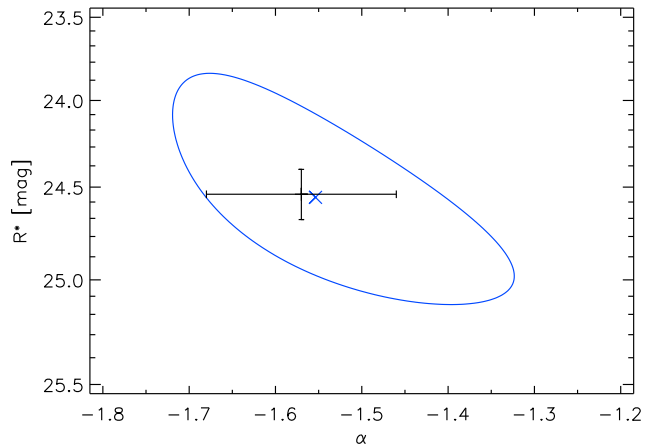}}
   \caption{The 1$\sigma$ confidence contour for the characteristic
            magnitude ($R^*$) and faint end slope ($\alpha$) fitted under
            the assumption that GRBs trace star formation. The $\times$ 
            indicates the best fit. The point with errorbars marks the 
            $R^*$ and $\alpha$ range for LBGs \citep{adelberger00}.}
\label{maxlik.fig}
\end{figure}

\section{Summary}
We have carried out a narrow-band Ly$\alpha$ imaging of three GRB host
galaxies and their environments. The GRB\,021004 host ($z = 2.335$) is a 
strong Ly$\alpha$ emitter with a restframe EW of $68^{+12}_{-11}$\,\AA. 
We no not detect the hosts of GRB\,030226 ($z = 1.986$) and GRB\,020124 
($z = 3.198$) in Ly$\alpha$; the $5 \sigma$ upper limits are 
$2.4 \times 10^{-17}$\,erg\,s$^{-1}$\,cm$^{-2}$ and 
$1.5 \times 10^{-17}$\,erg\,s$^{-1}$\,cm$^{-2}$, respectively. We
find a total of 12 LEGOs in the fields of GRB\,021004 and GRB\,030226.
\par
We have also assembled a list of all GRB hosts at \mbox{$z \ga 2$} reported
in the literature. The distribution of the host magnitudes is consistent 
with the LF of LBGs at $z = 3$. If the LBG restframe UV continuum is 
proportional to the SFR, this suggests that the hypothesis of GRBs as 
star-formation tracers is compatible with current observations. In addition, 
when applying the mass-metallicity relation \citep{tremonti}, this would 
imply that low-mass galaxies dominate the integrated star-formation density 
at all epochs.
\par
To clarify the situation we need a complete and unbiased sample of GRB 
host galaxies with measured redshifts in a well defined redshift
range. This is necessary in order to exclude any remaining uncertainty 
on whether the current sample is biased against dusty host galaxies. The
currently operating \emph{Swift} satellite \citep{swift} offers a unique 
possibility to secure such a sample. We also need an independent gauge of 
the SFR distribution function in the same well defined redshift range. One
suitable redshift range is $z < 0.2$ where large redshift surveys, such as
the Sloan Digital Sky Survey \citep[SDSS, e.g.][]{sdss} and the 2dF 
Galaxy Redshift Survey \citep[2dFGRS, e.g.][]{2dF}, have provided an 
extensive census and where it is possible to 
firmly constrain the amount of obscured star formation. Another 
possible redshift range is $2 < z < 4$, where a plenitude of starburst 
selection techniques are available, e.g. LBGs, Ly$\alpha$ galaxies, 
sub-mm galaxies and distant red galaxies, and where the peak of the 
cosmic star-formation density appears to have been located.

\section*{Acknowledgments}
We thank Stephanie Courty for an intense and enlightening discussion on
(specific) star formation, and the anonymous referee for a careful 
reading and useful corrections. PJ, GB and EHG gratefully acknowledge support 
from a special grant from the Icelandic Research Council. JPUF acknowledges 
support from the Carlsberg foundation. BLJ acknowledges support from the 
Instrument Centre for Danish Astrophysics (IDA) and the Nordic Optical 
Telescope (NOT). JG acknowledges the support of a Ram\'on y Cajal Fellowship 
from the Spanish Ministry of Education and Science. The research of JG is 
supported by the Spanish Ministry of Science and Education through 
programmes ESP2002-04124-C03-01 and AYA2004-01515. This work was supported 
by the Danish Natural Science Research Council. The authors acknowledge 
benefits from collaboration within the EU FP5 Research Training Network 
``Gamma-Ray Bursts: An Enigma and a Tool". We are also grateful to the 
Nordic Optical Telescope Scientific Association (NOTSA) and the Science 
Institute, University of Iceland for financial support.

\appendix
\section{Maximum Likelihood Analysis}
\label{fit.app}
The Schechter luminosity function (LF) is given by:
\[
\Phi (L) \mathrm{d}L = \Phi^{*} \left( \frac{L}{L^{*}} \right)^{\alpha}
\exp \left( - \frac{L}{L^{*}} \right) \mathrm{d} 
\left( \frac{L}{L^{*}} \right) ,
\]
where $\Phi^{*}$ is a number per unit volume and $L^{*}$ is a 
characteristic luminosity (with an equivalent characteristic
magnitude) at which the LF exhibits a rapid change
in the slope in the $(\log \Phi, \log L)$-plane. The dimensionless
parameter $\alpha$ gives the slope of the LF in
the $(\log \Phi, \log L)$-plane when $L \ll L^{*}$.
\par
If the probability of a GRB occurring in a galaxy is proportional to
its UV luminosity, then the probability to find a GRB in a galaxy
with UV luminosity $L$ is proportional to $L \times \Phi(L)$. 
Normalising this probability density we find:
\[
p(\alpha,L^{*};L) = \frac{L^{\alpha+1}\exp(-L/L^{*})}
{(L^{*})^{\alpha+2} \, \Gamma(\alpha+2)} ,
\]
where $\Gamma$ is the gamma-function. In the discussion we use the
LBG restframe UV LF and compare with the sample of $z>2$ GRB hosts. 

Given the observed sample, we use the following likelihood function 
($LH$) for $L^{*}$ and $\alpha$:
\[
LH (\alpha, L^{*}) = \prod_{i=1}^{N}p(\alpha,L^{*};L_i) \times 
\prod_{j=1}^{M}
\int_{0}^{L_j}p(\alpha,L^{*};L) \mathrm{d}L \, ,
\]
where $N = 7$ is the number of GRB hosts with a detected luminosity ($L_i$)
and $M = 6$ is the number of GRB hosts with a luminosity upper limit ($L_j$). 
The maximum likelihood corresponds to an observed $R$-band magnitude of 
24.8 and a faint end slope of $\alpha=-1.53$ (Fig.~\ref{maxlik.fig}). 
The 1$\sigma$ error contour in the ($R^*$, $\alpha$)-plane is found 
from the condition:
\[
\log_{\mathrm{e}} \frac{LH(\alpha, R^*)}{LH_{\mathrm{max}}} = -1/2 \, .
\]

\label{lastpage}

\end{document}